\documentclass[twocolumn,showpacs,preprintnumbers,amsmath,amssymb]{revtex4}
\usepackage{tabularx,graphicx}\begin{document}
\newcommand{\beq}{\begin{equation}}
\newcommand{\eeq}{\end{equation}}
\newcommand{\beqn}{\begin{eqnarray}}
\newcommand{\eeqn}{\end{eqnarray}}
\newcommand{\bmath}{\begin{subequations}}
\newcommand{\emath}{\end{subequations}}
\title{Reply to ``Comment on `Charge expulsion and electric field in superconductors' ''}
\author{J. E. Hirsch }
\address{Department of Physics, University of California, San Diego\\
La Jolla, CA 92093-0319}
 
\date{\today} 
\begin{abstract} 
I argue that the validity of the new electrodynamic equations for superconductors proposed in my paper can and should be decided by experiment.  Furthermore I show that BCS theory is ambiguous in its prediction of screening of longitudinal
electric fields and hence cannot be used to decide between the validity of my theory versus
 the conventional theory. 
I also give a physical argument for why screening of electric fields by superconducting electrons should be expected to be less efficient than that by normal electrons, as 
predicted by the new equations.

\end{abstract}
\maketitle 

In my recent work\cite{charge,electro}, a new relation between charge density $\rho$ and electric potential
$\phi$ in superconductors was proposed
\beq
\rho(\vec{r})-\rho_0=-\frac{1}{4\pi\lambda_L^2}[\phi(\vec{r})-\phi_0(\vec{r})]
\eeq
where $\lambda_L$ is the London penetration depth, $\rho_0$ is a positive
constant, and $\phi_0(\vec{r})$ is the electrostatic potential generated by charge
density $\rho_0$. In this comment\cite{comment} it is argued that the correct relation is instead
 \beq
\rho(\vec{r}) =-\frac{1}{4\pi\lambda_{TF}^2}\phi(\vec{r}) 
\eeq
with $\lambda_{TF}$ the Thomas Fermi screening length, and hence that Eq. (1) 
is incorrect.

There are two differences between Eq. (1) and Eq. (2). First, Eq. (1) together with
Poisson's equation predicts the existence of an excess of negative charge within
a London penetration depth of the surface of a superconductor, and an
excess of positive charge (of density $\rho_0$) in the interior. As a consequence,
a 'spontaneous' electrostatic field is predicted to exist inside superconductors and, 
if the sample is non-spherical, also outside\cite{ellipse}. Instead, Eq. (2) predicts
that no charge inhomogeneity and hence no electrostatic field exists in the
absence of applied external field. The validity of either of these predictions can be tested by experiment
but this has not been done yet. 

Second, Eq. (1) predicts that externally applied electrostatic fields will penetrate
the superconductor a distance $\lambda_L$, which is typically at least
two orders of magnitude larger than the normal state Thomas Fermi penetration 
length $\lambda_{TF}$ which Eq. (2) predicts. In 1936 H. London performed
an experiment which in principle should have been able to discriminate between
these predictions\cite{london}, namely he attempted to measure a change in
the capacitance of a capacitor with superconducting plates when it goes
superconducting. He found no change, which would
appear to support Eq. (2) over Eq. (1). However, I propose Eq. (1) to describe only the
superfluid response, and at the temperature at which   H. London's experiment was performed
($T/T_c\sim 0.4$) , it would have
been masked by the normal quasiparticle response, which is of the form Eq. (2): the expected increase
in the electric penetration length at that temperature (due to the decrease in the density of normal fluid carriers)
is less than a factor of two of the 
normal metal value, far below London's experimental sensitivity which was of order $20 \AA$.  The
experiment has never been repeated at lower temperatures, hence this second aspect of Eq. (1)
also remains experimentally untested.

The comment objects to Eq. (1) on theoretical grounds. First it argues that 
``Eq. (1) cannot be accepted, because it is against the general understanding that the
charge screening length is still given by $\lambda_{TF}$ in the superconducting
state''. To counter this qualitative statement I give the following qualitative argument: mobile electrons in
normal metals screen because they are driven by the electric field to the point where they have minimum
{\it potential energy}. Instead for superfluid electrons {\it both potential and  kinetic energy} play a role, and the point of minimum total 
energy will in general be different from that of minimum potential energy. To put it another way, when
a superfluid electron moves to a region of lower potential energy it gains kinetic energy; since it cannot lose
this kinetic energy by inelastic scattering as a normal electron would, it will 'overshoot' the point of minimum
potential energy leaving the electric field unscreened.  Thus superfluid electrons should be expected to be less efficient
at screening longitudinal electric fields than normal electrons.

Second, the comment argues for the validity of Eq. (2) over Eq. (1) based on
BCS theory. This assumes the validity of BCS theory to describe superconductivity,
an assumption that my paper doesn't make. Contrary to the situation in 1969\cite{parks}
when it was generally believed that BCS theory describes all superconductivity
in solids, it is currently generally agreed that at least  $some$ superconductors
(e.g. high $T_c$ cuprates) are $not$ described by conventional BCS theory. Hence it is
not inconceivable that the electrodynamics of superconductors deduced from
BCS theory may not describe the electrodynamics of real superconductors.

Furthermore, the calculation described in the comment within BCS theory that 
reaches the result Eq. (2) is done {\it within a particular gauge}, where the
magnetic vector potential is set to zero. There is no a priori reason to choose
that gauge over any other gauge, and in any other gauge the BCS calculation
will yield a result different from Eq. (2). Related to this, an alternative way to do the
calculation  within BCS theory is to compute the
current density $\vec{J}$ in the presence of a space- and time-dependent potential
with Fourier component $\phi(\vec{q},\omega)$, which yields\cite{rick}
\beq
\vec{q}\cdot \vec{J}(\vec{q},\omega)=\frac{n_se^2}{m}\frac{q^2}{\omega}\phi(\vec{q},\omega)
\eeq
for small $q$ and $\omega$, with $n_s$ the superfluid density. Next we extract the charge density from the continuity
equation
\beq
\vec{\nabla}\cdot \vec{J}=-\frac{\partial \rho}{\partial t}, 
\eeq
and obtain 
\beq
\rho=-\frac{n_se^2}{m}\frac{q^2}{\omega^2}\phi .
\eeq
Assuming that both $\lambda_L$ and $\lambda_{TF}$ are given by their free electron expressions
\bmath
\beq
\frac{1}{\lambda_L^2}=\frac{4\pi n_s e^2}{mc^2}
\eeq
\beq
\frac{1}{\lambda_{TF}^2}=\frac{6\pi n_s e^2}{\epsilon_F}
\eeq
\emath
with $\epsilon_F$ the Fermi energy, Eq. (5) will yield my result Eq. (1) (for the particular case $\rho_0=0$) if the limit
$(q,\omega)\rightarrow(0,0)$ is taken such that $\omega/q=c$, and will yield the result of the comment
Eq. (2) if the limit is taken with $\omega/q=v_F/\sqrt{3}$, with $v_F$ the Fermi velocity, and will yield   any other 
result depending on how the limit is taken. This illustrates the ambiguity of the BCS approach and that {\it it
cannot be used} to decide between the validity of Eqs. (1) and (2).

I note that Eq.  (1) (for $\rho_0=0$) and Eq. (2) are both of the form
\beq
\rho(\vec{r}) =-en_s(\frac{e\phi(\vec{r})}{E_{sc}}) 
\eeq
which says that the fraction of superfluid density $n_s$ that gives rise to the induced charge
density is given by the ratio of the electrostatic energy to an energy scale $E_{sc}$. For Eq. (2) this
energy scale is $E_{sc}\sim \epsilon_F$, the Fermi energy, while for Eq. (1) it is $E_{sc}=mc^2$,
the rest energy of the superfluid condensate carriers. Furthermore in the classical theory of the
condensed charged Bose gas\cite{bose}, this energy scale is given by $E_{sc}\sim \hbar\omega_p$,
with $\omega_p$ the plasma frequency. These are three different possible answers
 that give rise to three different screening lengths, and certainly 
without assuming the validity of a given microscopic theory
it is not possible to decide between these  answers on theoretical grounds. And even assuming the
validity of BCS theory the answer is ambiguous as discussed above.

The electrodynamics that I propose is simpler than the one advocated in the comment, in that it relies on
the Lorenz gauge rather than the unnatural gauge given by Eq. (6)  of the comment. The dispersion
relation for the longitudinal and transverse modes is identical in my theory\cite{electro}, while it is very different in the
theory advocated in the comment. Furthermore the electrodynamics I propose is relativistically covariant while
the one  in the comment is not. Thus I argue that my theory is more natural than the
conventional theory advocated  in the comment and for that reason more likely to be correct. Nevertheless, the
validity or invalidity of it should be decided by experiment  and not by
theoretical prejudice. In addition to the experimental tests discussed above, measurement of the dispersion relation
for longitudinal plasmon modes should be able to differentiate between both theories, but such measurement  in the superconducting state has not
been done to date.

\end{document}